\documentclass[preprint,pra,10pt,a4paper,twocolumn,final,tightenlines,showpacs,notitlepage]{revtex4}

\input{tcilatex}

\begin{document}

\title{Classical information capacities of some single qubit quantum noisy
channels}
\author{Liang Xian-Ting}
\address{$^{1}$Department of Physics and Institute of Mathematics,\\
Huaihua College, Huaihua, Hunan 418008, China\\
$^{2}$Department of Material Science and Engineering, University \\
of Science and Technology of China, Hefei, Anhui 230026, China}
\pacs{03.67.H}
\keywords{Classical information capacity; Quantum noisy channel; HSW theorem}

\begin{abstract}
By using the Holevo-Schumacher-Westmoreland (HSW) theorem and through
solving eigenvalues of states out from the quantum noisy channels directly,
or with the help of the\ Bloch sphere representation, or Stokes
parametrization representation, we investigate the classical information
capacities of some well-known quantum noisy channels.
\end{abstract}

\maketitle

\section*{Introduction}

Calculating the information capacities of quantum noisy channels is an
important task for studying quantum communication \cite{Holevo/9809023}. It
has attracted much interest and many methods and results are proposed and
obtained \cite{NandCbook}. This interest was mainly stimulated by interplay
between quantum communication theory and quantum information ideas related
to more recent development in quantum computing and quantum communication.
Unlike classical channels, which are adequately characterized by a single
capacity, a quantum channel has several distinct capacities. They include
(1) \emph{classical capacity} $C_{pp}$, for transmitting classical
information by encoding it with quantum product states and decoding it with
individual measurement \cite{Kholevo73} \cite{Fuchs96}; (2) \emph{classical
capacity} $C_{pe}$, for transmitting classical information by encoding it
with quantum product states and decoding it with collective measurement \cite%
{Holevo98} \cite{Hausladen96} \cite{Schumacher97}. Similarly, there are
another two capacities denoted by $C_{ep}$ and $C_{ee}$; (3) \emph{%
entanglement-assisted classical capacity }\cite{Holevo0106075} $C_{ea}$
which describes the capacity of transmitting intact quantum states by the
help of prior entanglement between the sender and the receiver; (4) \emph{%
quantum capacity }$Q,$ a supremum of \emph{coherent information} which is
the correspondence of mutual information in classical information theory %
\cite{Baruum9702049} \cite{Schumacher96}; and (5) \emph{classical assisted
quantum capacity} $Q_{c}$ \cite{Bennett99}. In general, it is difficult to
calculate these capacities by their definitions. Scientists discovered some
clear expressions which may simplify the calculation but there are still
involving some technical problems. So for calculating the capacities
distinctly some special methods are developed in last years. By using some
special methods we have investigated the entanglement-assisted classical
information capacities of some single qubit quantum noisy channels \cite%
{LF02}. In this paper our investigation focus on classical information
capacities $C_{pe}$ for some well-known quantum noisy channels. In
subsection 2.1 we will investigate the capacities of depolarizing and
erasure quantum noisy channels. The capacities of these two kinds of
channels can be calculated by solving eigenvalues of output states directly.
In subsection 2.2 we will calculate the capacities of serval well-known
quantum noisy channels with the help of Bloch sphere representation of qubit
quantum state. In subsection 2.3 we use the Stokes parametrization
representation of qubit quantum state to investigate the capacities of
amplitude damping channel and splaying channel. A brief conclusion will
close this paper in the last section.

\section{Calculating the capacities of quantum noisy channels}

Interactions with the environment are the fundamental source of noise in
both classical and quantum systems. It is often not easy to find exact
models for the environment or the system-environment interaction. However,
some quantum noisy models, for example, depolarizing channel, phase damping
channel, two-Pauli channel, amplitude damping channel etc. \cite{Uhlmann01}
can attain a high degree of accuracy in modeling of noise in circuits of
quantum computation and quantum communication \cite{Liang02}. So calculating
the capacities of quantum noisy channels is a significative work. We start
our research with reviewing some concepts as follows.

$\star $ Operator sum representation: Every completely positive
trace-preserving map $\varepsilon $ can\ be regarded as a channel which can
be represented (non-uniquely) in the Kraus form%
\begin{equation}
\rho ^{\prime }=\varepsilon \left( \rho \right) =\tsum E_{k}^{\dagger }\rho
E_{k}  \label{eq1}
\end{equation}%
which is also called operator sum representation. Here, $E_{k}$ is the Kraus
operators and $\rho $ is the density matrix of input state and $\rho
^{\prime }$ is the density matrix of output state.

$\star $ Unital map: \cite{King01} If $\varepsilon $ map the identity
operator to itself i.e.%
\begin{equation}
\varepsilon \left( I\right) =I,  \label{eq2}
\end{equation}%
this map is a unital map.

$\star $ Bloch sphere representation: A density matrix of a qubit quantum
pure state can be expressed as%
\begin{equation}
\rho =\frac{1}{2}\left( 
\begin{array}{cc}
1+\cos \theta & e^{-i\varphi }\sin \theta \\ 
e^{i\varphi }\sin \theta & 1-\cos \theta%
\end{array}%
\right) .  \label{eq3}
\end{equation}%
This expression is called Bloch sphere representation of qubit quantum state.

$\star $ Holevo-Schumacher-Westmoreland (HSW) theorem: \cite{Holevo98}, \cite%
{Hausladen96}, \cite{Schumacher97} The classical information capacity
(encoding with pure product states and decoding with collective measurement)
of quantum channel is given by%
\begin{equation}
C_{pe}=\sup_{\left\{ p_{j},\rho _{j}\right\} }\left(
S(\tsum_{j}p_{j}\varepsilon \left( \rho _{j}\right) )-\tsum_{j}p_{j}S\left(
\varepsilon \left( \rho _{j}\right) \right) \right) .  \label{eq4}
\end{equation}%
Here, $S(\tau )=-\tau \log \left( \tau \right) $ ( where and throughout this
paper the logarithms are taken to base two) denotes the von Neumann entropy, 
$p_{j}$ is the probability of state $\rho _{j}$ in ensemble $\left\{
p_{j},\rho _{j}\right\} $.

$\star $ If a map is unital, then the classical information capacity $C_{pe}$
can be obtained with orthogonal input \cite{King01}.

By using above definitions, theorem and proposition we now investigate the
capacities $C_{pe}$ for some well-known quantum noisy channels.

\subsection{By solving the eigenvalues of output states directly}

In this subsection we will investigate the classical information capacities $%
C_{pe}$ of depolarizing channel and the erasure channel through solving
eigenvalues of output states directly.

$\bullet $ Depolarizing channel: At first, we investigate the memoryless
depolarizing channel (memoryless is assumed throughout this paper for all
channels). Depolarizing channel is a important type of quantum noisy
channels. It models a decohering qubit that has particularly nice symmetry
properties. There are many practical quantum processes corresponding to this
model \cite{NandCbook}. Its Kraus operators are%
\begin{eqnarray}
E_{0}^{d} &=&\sqrt{1-\eta }I,E_{1}^{d}=\sqrt{\frac{\eta }{3}}\sigma
_{1},E_{2}^{d}=\sqrt{\frac{\eta }{3}}\sigma _{2},E_{3}^{d}=\sqrt{\frac{\eta 
}{3}}\sigma _{3}.  \nonumber \\
&&  \label{eq5}
\end{eqnarray}%
$\sigma _{i}$ $(i=1,2,3)$ denote the Pauli matrices; $I$ is the identity
matrix in the Hilbert space of $C^{2\times 2}$; $\eta $ is the error
occurring probability of quantum state passing through a depolarizing
channel. So when a quantum state $\rho $ is transmitted through this channel
the state becomes%
\begin{eqnarray}
\rho ^{\prime } &=&\varepsilon ^{d}\left( \rho \right)   \nonumber \\
&=&\left( 1-\eta \right) \rho +\frac{\eta }{3}\left( \sigma _{1}\rho \sigma
_{1}+\sigma _{2}\rho \sigma _{2}+\sigma _{3}\rho \sigma _{3}\right) ,
\label{eq6}
\end{eqnarray}%
where $\varepsilon ^{d}$ denotes a map of depolarizing channel (we use the
first or the first two or three letters of the names of the channels as the
superscripts to differentiate the map $\varepsilon ,$\ Kraus operators $E_{i}
$ and so on of different channels); we set $\rho $ is the input pure state
and $\rho ^{\prime }$ is the output state (in general, it is mixed state) of
input state $\rho $; This map $\varepsilon ^{d}$ is a unital map. From the
following relationship%
\begin{equation}
2I=\rho +\sigma _{1}\rho \sigma _{1}+\sigma _{2}\rho \sigma _{2}+\sigma
_{3}\rho \sigma _{3}  \label{eq7}
\end{equation}%
we can obtain the two output states $\rho _{1}^{\prime }$ and $\rho
_{2}^{\prime }$ of a pair of input encoding states $\rho _{1}$ and $\rho _{2}
$ as%
\begin{equation}
\rho _{1,2}^{\prime }=\left( 1-\frac{4\eta }{3}\right) \rho _{1,2}+\frac{%
2\eta }{3}I.  \label{eq8}
\end{equation}%
$\rho _{1}^{\prime }$ has the eigenvalues $\alpha _{1}^{d}=1-2\eta /3,$ $%
\alpha _{2}^{d}=2\eta /3,$ similarly $\rho _{2}^{\prime }$ has the
eigenvalues $\beta _{1}^{d}=1-2\eta /3,$ $\beta _{2}^{d}=2\eta /3$ (in the
following we always denote the eigenvalues of states $\rho _{1}^{\prime }$
with $\alpha _{i}$ and $\rho _{2}^{\prime }$ with $\beta _{i}$ $(i=1,2)$).
So $S\left( \varepsilon ^{d}\left( \rho _{j}\right) \right) =H\left( 2\eta
/3\right) $ which does not depend on $\rho _{j}$ at all. Here, $H\left( \tau
\right) $ denotes the binary entropy i.e. $H(\tau )=-\tau \log \tau -(1-\tau
)\log (1-\tau )$. Thus, $C_{pe}$ can be achieved by maximizing the entropy
of mixed state $\varrho ^{d}=\tsum\nolimits_{j}p_{j}\varepsilon ^{d}\left(
\rho _{j}\right) $ which may be done through a pair orthogonal input states
(we denote the mixed density matrix $\tsum\nolimits_{j}p_{j}\varepsilon
\left( \rho _{j}\right) $ always by $\varrho $ in this paper). For example,
we calculate it by simply choosing the $\left| 0\right\rangle $ and $\left|
1\right\rangle $ of single qubit, so we immediately obtain a eigenvalues of
mixed state $\varrho ^{d}$ as $\gamma _{1}^{d}=\gamma _{2}^{d}=1/2$ (we
always denote the eigenvalues of mixed state $\varrho
=\tsum\nolimits_{j}p_{j}\varepsilon \left( \rho _{j}\right) $ by $\gamma _{i}
$ ). Thus, we can obtain its capacity as%
\begin{equation}
C_{pe}^{d}=1-H\left( \frac{2\eta }{3}\right) .  \label{eq9}
\end{equation}

$\bullet $ Erasure channel: The capacity of another quantum noisy channel,
erasure channel can also be investigated as same as depolarizing channel %
\cite{GBP97}. When a quantum state is transmitted through this channel the
undisturbed probability is $1-\eta $. In case of an error, the quantum state
is replaced by $\left| \xi \right\rangle $ that is orthogonal to all quantum
states of the system. In another words, the error make the state out of its
original Hilbert space i.e. the information is erased with probability $\eta
.$ So when a quantum state $\rho \in \mathcal{H}=C^{2\times 2}$ pass through
this channel the state becomes%
\begin{eqnarray}
\rho ^{\prime } &=&\varepsilon ^{e}\left( \rho \right) =\left( 1-\eta
\right) \rho +\eta \left| \xi \right\rangle \left\langle \xi \right| ,\left|
\xi \right\rangle \notin \mathcal{H}  \nonumber \\
&&  \label{eq10}
\end{eqnarray}%
The capacity can be easily calculated. It is $1-\eta $ \cite{BDS97},\cite%
{Keyl0202112}.

\subsection{By help with Bloch sphere representation}

The classical information capacities of some other quantum noisy channels
may not be calculated as easier as above two channels. For example, the
capacities of phase damping channel, two Pauli channel, bit flip channel
etc. can not be calculated by solving the eigenvalues of the output states
directly. In this case, we find that it may be convenient by using Bloch
sphere representation of qubit quantum state. Analytically investigating the
classical information capacities by using the Bloch sphere representations
of qubit states is also restricted in some quantum noisy channels which are
expressed by unital maps, but by using this method and with the help of
numerical work we can widely investigate the classical information
capacities $C_{pe}$ almost for all of the quantum noisy channels. So this
method is very powerful.

Now let us calculate the classical information capacity $C_{pe}$ of phase
damping channel by use of Bloch sphere representation.

$\bullet $ Phase damping channel: It has the Kraus operators as \cite%
{Preskill}%
\begin{eqnarray}
E_{0}^{p} &=&\sqrt{1-\eta }I,E_{1}^{p}=\sqrt{\eta }\left[ 
\begin{array}{cc}
1 & 0 \\ 
0 & 0%
\end{array}%
\right] ,E_{2}^{p}=\sqrt{\eta }\left[ 
\begin{array}{cc}
0 & 0 \\ 
0 & 1%
\end{array}%
\right] .  \nonumber \\
&&  \label{eq11}
\end{eqnarray}%
By using Eqs.(\ref{eq1}) and (\ref{eq11}) we can obtain the density matrix
after a state $\rho $ transmitting through this channel as%
\begin{equation}
\rho ^{\prime }=\varepsilon ^{p}(\rho )=\left( 1-\mu \right) \rho +\mu
\sigma _{3}\rho \sigma _{3},  \label{eq12}
\end{equation}%
where $\mu =\eta /2.$ It can be easily seen that this channel is unital we
can obtain its capacity by input a pair orthogonal states. By using the
Bloch presentation we have%
\begin{eqnarray}
\rho _{j}^{\prime } &=&\varepsilon ^{p}\left( \rho _{j}\right)  \nonumber \\
&=&\frac{1}{2}\left( 
\begin{array}{cc}
\left( 1+\cos \theta _{j}\right) & e^{-i\varphi }(1-2\mu )\sin \theta _{j}
\\ 
e^{i\varphi }(1-2\mu )\sin \theta _{j} & (1-\cos \theta _{j})%
\end{array}%
\right) ,  \nonumber \\
&&  \label{eq13}
\end{eqnarray}%
where $j=1,2.$ From above equation we known that when the prior
probabilities $p_{1}=p_{2}=\frac{1}{2},$ $S\left( \varrho ^{p}\right)
=S\left( \sum_{j}p_{j}\varepsilon ^{p}\left( \rho _{j}\right) \right) $ take
its maximum value $1bit$ at $\theta _{2}-\theta _{1}=\pi $ (orthogonal).
Because the eigenvalues of $\varepsilon ^{p}(\rho _{1})$ at $\theta _{1}=0,$
are%
\begin{eqnarray}
\alpha _{1}^{p} &=&\frac{1}{2}+\frac{1}{2}\sqrt{1-4\mu \left( 1-\mu \right)
\sin ^{2}\theta _{1}}=1,  \nonumber \\
\alpha _{2}^{p} &=&\frac{1}{2}-\frac{1}{2}\sqrt{1-4\mu \left( 1-\mu \right)
\sin ^{2}\theta _{1}}=0.  \label{eq14}
\end{eqnarray}%
Similarly, the eigenvalues of $\varepsilon ^{p}(\rho _{1})$ at $\theta
_{1}=\pi ,$ are%
\begin{equation}
\beta _{1}^{p}=1,\qquad \beta _{2}^{p}=0.  \label{eq15}
\end{equation}%
So the classical information capacity $C_{pe}$ of the phase damping channel
is $1bit$ which corresponds to the input encoding states $\left|
0\right\rangle ,$ and $\left| 1\right\rangle .$

Similarly, we can calculate the classical information capacities of bit
flip, bit-phase flip, and phase flip channels.

$\bullet $ Bit flip channel: The Kraus operators of bit flip channel are%
\begin{equation}
E_{0}^{bf}=\sqrt{1-\eta }I\mathbf{,\qquad }E_{1}^{bf}=\sqrt{\eta }\sigma
_{1}.  \label{eq16}
\end{equation}

$\bullet $ Bit-phase flip channel: It has Kraus operators as%
\begin{equation}
E_{0}^{bpf}=\sqrt{1-\eta }I\mathbf{,\qquad }E_{1}^{bpf}=\sqrt{\eta }\sigma
_{2}.  \label{eq17}
\end{equation}

$\bullet $ Phase flip channel: Its Kraus operators are 
\begin{equation}
E_{0}^{pf}=\sqrt{1-\eta }I\mathbf{,\qquad }E_{1}^{pf}=\sqrt{\eta }\sigma
_{3}.  \label{eq18}
\end{equation}%
The capacities of these channels are all $1bit$ which correspond to the
input encoding states $\frac{1}{\sqrt{2}}\left( \left| 0\right\rangle
+\left| 1\right\rangle \right) ,$ and $\frac{1}{\sqrt{2}}\left( \left|
0\right\rangle -\left| 1\right\rangle \right) $ $\left( \text{bit flip
channel}\right) ,$ $\frac{i}{\sqrt{2}}\left( \left| 0\right\rangle +\left|
1\right\rangle \right) ,$ and $\frac{i}{\sqrt{2}}\left( -\left|
0\right\rangle +\left| 1\right\rangle \right) $ $\left( \text{bit-phase flip
channel}\right) ,$ $\left| 0\right\rangle ,$ and $\left| 1\right\rangle $ $%
\left( \text{phase flip channel}\right) $. These results can also be
intuitively seen from the evolution of Bloch spheres in these quantum
channels, because their Bloch spheres have the same form as Bloch sphere in
phase damping channel except for their directions \cite{NandCbook}.

Now we investigate two-Pauli channel by using this method.

$\bullet $ Two-Pauli channel: The Kraus operators of two-Pauli channel are 
\begin{equation}
E_{0}^{t}=\sqrt{1-\eta }I\mathbf{,}E_{1}^{t}=\sqrt{\frac{\eta }{2}}\sigma
_{1},E_{2}^{t}=\sqrt{\frac{\eta }{2}}\sigma _{2},  \label{eq19}
\end{equation}%
From Eqs.(\ref{eq1}) and (\ref{eq19}) we have 
\begin{eqnarray}
\rho _{j}^{\prime } &=&\varepsilon ^{t}\left( \rho _{j}\right)  \nonumber \\
&=&\frac{1}{2}\left( 
\begin{array}{cc}
1+(1-2\eta )\cos \theta _{j} & e^{-i\varphi }\left( 1-\eta \right) \sin
\theta _{j} \\ 
e^{i\varphi }\left( 1-\eta \right) \sin \theta _{j} & 1-(1-2\eta )\cos
\theta _{j}%
\end{array}%
\right) ,  \nonumber \\
&&  \label{eq20}
\end{eqnarray}%
From above equation we known $S\left( \varrho ^{t}\right) =S\left(
\tsum\nolimits_{j}p_{j}\varepsilon ^{t}\left( \rho _{j}\right) \right) $
takes its maximum value $1bit$ at $\theta _{1}-\theta _{2}=\pi $ when the
prior probabilities $p_{1}=p_{2}=\frac{1}{2}$. The eigenvalues of $%
\varepsilon ^{t}(\rho _{1})$ are 
\begin{eqnarray}
\alpha _{1,2}^{t} &=&\frac{1\pm \sqrt{1-4\eta \left( 1-\eta \right) +\eta
\left( 2-3\eta \right) \sin ^{2}\theta _{1}}}{2},  \nonumber \\
&&  \label{eq21}
\end{eqnarray}%
Similarly, the eigenvalues of $\varepsilon ^{t}(\rho _{2})$ are 
\begin{eqnarray}
\beta _{1,2}^{t} &=&\frac{1\mp \sqrt{1-4\eta \left( 1-\eta \right) +\eta
\left( 2-3\eta \right) \sin ^{2}\theta _{2}}}{2}.  \nonumber \\
&&  \label{eq22}
\end{eqnarray}%
So, when $0<\eta <\frac{2}{3},$ set $\theta _{1}=\frac{\pi }{2}$ and $\theta
_{2}=-\frac{\pi }{2},$ $\alpha _{1,2}^{t}=\frac{1}{2}\pm \frac{1}{2}\left(
1-\eta \right) ,$ $\beta _{1,2}^{t}=\frac{1}{2}\mp \frac{1}{2}\left( 1-\eta
\right) $ and when $\frac{2}{3}\leq \eta <1,$ $\alpha _{1,2}^{t}=\frac{1}{2}%
\pm \frac{1}{2}\left( 2\eta -1\right) $, $\beta _{1,2}^{t}=\frac{1}{2}\mp 
\frac{1}{2}\left( 2\eta -1\right) $ which correspond to the minimum value of
von Neumann entropies of $\varepsilon ^{t}(\rho _{j}^{\prime }).$ So the
capacity of this channel is%
\begin{equation}
C_{pe}^{t}=\left\{ 
\begin{array}{c}
1-H\left( \frac{1}{2}\eta \right) ,\qquad 0<\eta <\frac{2}{3}, \\ 
1-H\left( \eta \right) ,\qquad \frac{2}{3}\leq \eta <1.%
\end{array}%
\right.  \label{eq23}
\end{equation}%
It has been shown that the Bloch sphere representation is a powerful tool
for analyzing the classical information capacities of quantum noisy channels.

\subsection{By help with Stokes parametrization representation}

As mention above, analytically solving the capacities with Bloch sphere
representation is not perfect effective for non-unital quantum channels. At
this rate, Stokes parametrization representation \cite{King01} may help us.
In this subsection we will use Stokes parametrization representation to
investigate the capacities of amplitude-damping channel \cite{Preskill} and
``spraying'' channel \cite{Fuchs97}. Before our furthermore research we
review several concepts that will be used in the following.

$\star $ The identity and Pauli matrices form a basis for $C^{2\times 2}$ so
that any quantum state $\rho $ of qubit can be written as \cite{King01}%
\begin{equation}
\rho =\frac{1}{2}\left( I+\vec{w}\cdot \vec{\sigma}\right) ,  \label{eq24}
\end{equation}%
where $\vec{w}\ $is real and $\vec{w}\mathbf{\in }C^{3}$, $\vec{\sigma}%
=\left( \sigma _{1},\sigma _{2},\sigma _{3}\right) ^{\tau }$ ($\tau $
denotes the transpose of matrix).

$\star $ Any quantum state $\rho $ pass through a quantum noisy channel $%
\varepsilon $ becomes \cite{King01}%
\begin{eqnarray}
\rho ^{\prime } &=&\varepsilon \left( \frac{1}{2}\left( I+\vec{w}\cdot \vec{%
\sigma}\right) \right)   \nonumber \\
&=&\frac{1}{2}\left( I+\left( \vec{t}+T\vec{w}\right) \cdot \vec{\sigma}%
\right) ,  \label{eq25}
\end{eqnarray}%
where 
\begin{equation}
T\mathbf{=}\left( 
\begin{array}{ccc}
\chi _{1} & 0 & 0 \\ 
0 & \chi _{2} & 0 \\ 
0 & 0 & \chi _{3}%
\end{array}%
\right) ,\qquad \vec{t}=\left( 
\begin{array}{c}
t_{1} \\ 
t_{2} \\ 
t_{3}%
\end{array}%
\right) .  \label{eq26}
\end{equation}%
The unital maps correspond to $\vec{t}=0.$

$\bullet $ Amplitude damping channel: From the Kraus operators of amplitude
damping channel 
\begin{equation}
E_{0}^{a}=\left( 
\begin{array}{cc}
1 & 0 \\ 
0 & \sqrt{1-\eta }%
\end{array}%
\right) ,\qquad E_{1}^{a}=\left( 
\begin{array}{cc}
0 & \sqrt{\eta } \\ 
0 & 0%
\end{array}%
\right)   \label{eq27}
\end{equation}%
we can obtain that 
\begin{equation}
\varepsilon ^{a}\left( I\right) =\left( 
\begin{array}{cc}
1+\eta  & 0 \\ 
0 & 1-\eta 
\end{array}%
\right)   \label{eq28}
\end{equation}%
which shows that it is a non-unital channel. From its Kraus operators we can
know that the $T$ and $\vec{t}$ correspond to amplitude damping channel are%
\begin{equation}
T^{a}\mathbf{=}\left( 
\begin{array}{ccc}
\sqrt{1-\eta } & 0 & 0 \\ 
0 & \sqrt{1-\eta } & 0 \\ 
0 & 0 & 1-\eta 
\end{array}%
\right) ,\quad \vec{t}^{a}=\left( 
\begin{array}{c}
0 \\ 
0 \\ 
\eta 
\end{array}%
\right) ,  \label{eq29}
\end{equation}%
namely, $\chi _{1}=\chi _{2}=\sqrt{1-\eta },$ $\chi _{3}=1-\eta ,$ $%
t_{1}=t_{2}=0,$ $t_{3}=\eta .$ From Eq.(\ref{eq25}) we can obtain the output
state of input state $\rho =\frac{1}{2}\left( I+\vec{w}\cdot \vec{\sigma}%
\right) $ as 
\begin{eqnarray}
\rho ^{\prime } &=&\varepsilon ^{a}\left( \frac{1}{2}\left( I+\vec{w}\cdot 
\vec{\sigma}\right) \right)   \nonumber \\
&=&\frac{1}{2}\left( I+\sqrt{1-\eta }w_{1}\sigma _{1}+\sqrt{1-\eta }%
w_{2}\sigma _{2}+\sigma _{3}\right) ,  \nonumber \\
&&  \label{eq30}
\end{eqnarray}%
Due to $\chi _{1}=\chi _{2}>\chi _{3},$ and for $0\leq \eta \leq 1\ $in
order to calculate the capacity we set $w_{1}=1,$ $w_{2}=w_{3}=0.$ Thus, we
have 
\begin{equation}
\rho ^{\prime }=\frac{1}{2}\left( 
\begin{array}{cc}
1+\eta  & \chi _{1} \\ 
\chi _{1} & 1-\eta 
\end{array}%
\right) .  \label{eq31}
\end{equation}%
Because this channel is non-unital we do not know if orthogonal inputs can
give the classical information capacity or not and we also do not know in
advance what prior probabilities of input states give the maximum output
information. In the following we consider the case of only a pair input
encoding states. We suppose one of the input states denoted by $\rho _{1}=%
\frac{1}{2}\left( I+\vec{w}\cdot \vec{\sigma}\right) $ and the other denoted
by $\rho _{2}=\frac{1}{2}\left( I+\vec{w}^{\prime }\cdot \vec{\sigma}\right) 
$, thus after the state $\rho _{1}$ transmitting though this channel we can
obtain its output state as $\rho _{1}^{\prime }=\varepsilon ^{a}\left( \frac{%
1}{2}\left( I+\vec{w}\cdot \vec{\sigma}\right) \right) $ and state $\rho _{2}
$ transmitting though this channel we can obtain its output state as $\rho
_{2}^{\prime }=\varepsilon ^{a}\left( \frac{1}{2}\left( I+\vec{w}^{\prime
}\cdot \vec{\sigma}\right) \right) .$ As the Bloch sphere of the amplitude
damping channel is symmetrical about its $z$ axes i.e. the axes of $\theta =0
$ (the meaning of $\theta $ we refer the readers to Eq.(\ref{eq3})) $\vec{w}%
^{\prime }$ would be obtained by rotating $\vec{w}$ with $\psi $ angle
around the axes $z$, namely, $\vec{w}^{\prime }=R^{a}\vec{w},$ where%
\begin{equation}
R^{a}=\left( 
\begin{array}{ccc}
\cos \psi  & \sin \psi  & 0 \\ 
-\sin \psi  & \cos \psi  & 0 \\ 
0 & 0 & 1%
\end{array}%
\right) .  \label{eq32}
\end{equation}%
So we have%
\begin{equation}
\rho _{1}^{\prime }=\frac{1}{2}\left( 
\begin{array}{cc}
1+\eta  & \chi _{1} \\ 
\chi _{1} & 1-\eta 
\end{array}%
\right) ,  \label{eq33}
\end{equation}%
the eigenvalues of which are 
\begin{equation}
\alpha _{1,2}^{a}=\frac{1}{2}\pm \frac{1}{2}\sqrt{1-\eta +\eta ^{2}}.
\label{eq34}
\end{equation}%
\begin{eqnarray}
\rho _{2}^{\prime } &=&\varepsilon ^{a}\left( \frac{1}{2}\left( I+\vec{w}%
^{\prime }\cdot \vec{\sigma}\right) \right)   \nonumber \\
&=&\frac{1}{2}\left( 
\begin{array}{cc}
1+\eta  & \chi _{1}\cos \psi +i\chi _{2}\sin \psi  \\ 
\chi _{1}\cos \psi -i\chi _{2}\sin \psi  & 1-\eta 
\end{array}%
\right) .  \nonumber \\
&&  \label{eq35}
\end{eqnarray}%
The eigenvalues of $\rho _{2}^{\prime }$ are also%
\begin{equation}
\beta _{1,2}^{a}=\frac{1}{2}\pm \frac{1}{2}\sqrt{1-\eta +\eta ^{2}}.
\label{eq36}
\end{equation}%
It is shown that the different output states have the same von Neumann
entropy, namely,%
\begin{equation}
\tsum_{j}p_{j}S(\rho _{j}^{\prime })=H(\frac{1}{2}-\frac{1}{2}\sqrt{1-\eta
+\eta ^{2}}),  \label{eq37}
\end{equation}%
which is nether correlative to prior probabilities nor to the angle $\psi .$
So in order to maximize the output information, i.e. obtain the capacity of
this channel we only maximize $S\left( \varrho ^{a}\right) =S\left(
\sum_{j}p_{j}\varepsilon ^{a}\left( \rho _{j}\right) \right) $. We set $%
S\left( \varrho ^{a}\right) $ gets its maximum when $p_{1}=1-\tau $ and $%
p_{2}=\tau ,$ thus,%
\begin{eqnarray}
\varrho ^{a} &=&\sum_{i=1}^{2}p_{i}\rho _{i}^{\prime }=\left( 1-\tau \right)
\rho _{1}^{\prime }+\tau \rho _{2}^{\prime }  \nonumber \\
&=&\frac{1}{2}\left( 
\begin{array}{cc}
1+\eta  & \xi  \\ 
\xi  & 1-\eta 
\end{array}%
\right) .  \label{eq38}
\end{eqnarray}%
where $\xi =\chi _{1}\left( 1-\tau +\tau \cos \psi -i\sin \psi \right) .$
The eigenvalues of $\varrho ^{a}$ are 
\begin{equation}
\gamma _{1,2}^{a}=\frac{1\pm \sqrt{\eta ^{2}+\chi _{1}^{2}\left( 1-A(1-\cos
\psi \right) }}{2},  \label{eq39}
\end{equation}%
where $A=\tau \left( 1-\tau \right) ,$ $0<A<1.$ When we take $\psi =\pi $
and $\tau =\frac{1}{2},$ we obtain the maximum value of $S\left( \varrho
^{a}\right) =S\left( \sum_{j}p_{j}\varepsilon ^{a}\left( \rho _{j}\right)
\right) =H(\frac{1-\eta }{2}).$ Thus, the classical information capacity $%
C_{pe}$ of amplitude channel is%
\begin{equation}
C_{pe}^{a}=H(\frac{1-\eta }{2})-H(\frac{1-\sqrt{1-\eta +\eta ^{2}}}{2}).
\label{eq40}
\end{equation}%
In the following we will use this kind method to calculate the classical
information capacity of ``splaying'' channel.

$\bullet $ ``splaying'' channel: The $T$ and $\vec{t}$ correspond to this
channel are%
\begin{equation}
T^{s}=\left( 
\begin{array}{ccc}
\frac{1}{\sqrt{3}} & 0 & 0 \\ 
0 & 0 & 0 \\ 
0 & 0 & \frac{1}{3}%
\end{array}%
\right) ,\quad \vec{t}^{s}=\left( 
\begin{array}{c}
0 \\ 
0 \\ 
\frac{1}{3}%
\end{array}%
\right)   \label{eq41}
\end{equation}%
So the output state of input state $\rho _{1}=\frac{1}{2}\left( I+\vec{w}%
\cdot \vec{\sigma}\right) $ is 
\begin{eqnarray}
\rho _{1}^{\prime } &=&\varepsilon ^{s}\left( \frac{1}{2}\left( I+\vec{w}%
\cdot \vec{\sigma}\right) \right)   \nonumber \\
&=&\frac{1}{2}\left[ I+\chi _{1}w_{1}\sigma _{1}+\left( t_{3}+\chi
_{3}w_{3}\right) \sigma _{3}\right]   \label{eq42}
\end{eqnarray}%
Due to $\chi _{2}=0,$ $\chi _{1}>\chi _{3}$, in order to calculate the
capacity we set $w_{1}=1,$ $w_{2}=w_{3}=0.$ Thus, we have%
\begin{equation}
\rho _{1}^{\prime }=\left( 
\begin{array}{cc}
\frac{2}{3} & \frac{1}{2\sqrt{3}} \\ 
\frac{1}{2\sqrt{3}} & \frac{1}{3}%
\end{array}%
\right)   \label{eq43}
\end{equation}%
which has eigenvalues%
\begin{equation}
\alpha _{1}^{s}=\frac{5}{6},\qquad \alpha _{2}^{s}=\frac{1}{6}.  \label{eq44}
\end{equation}%
Now we hope to find another state by which and $\rho _{1}$ construct a
ensemble $\left\{ p_{j},\rho _{j}\right\} $ that make the splaying channel
had maximum output information. Because of $\chi _{2}=0,$ we can generally
obtain the state $\rho _{2}$ from $\rho _{1}$ by rotating $\vec{w}$ with $%
\psi ^{\prime }$ angle, namely, $\vec{w}^{\prime }=R^{s}\vec{w},$ where 
\begin{equation}
R^{s}=\left( 
\begin{array}{ccc}
\cos \psi ^{\prime } & 0 & \sin \psi ^{\prime } \\ 
0 & 1 & 0 \\ 
-\sin \psi ^{\prime } & 0 & \cos \psi ^{\prime }%
\end{array}%
\right) .  \label{eq45}
\end{equation}%
So the output state $\rho _{2}^{\prime }$ is 
\begin{eqnarray}
\rho _{2}^{\prime } &=&\varepsilon ^{s}\left( \frac{1}{2}\left( I+R^{\prime }%
\vec{w}\cdot \vec{\sigma}\right) \right)   \nonumber \\
&=&\frac{1}{2}\left[ I+\chi _{1}\sigma _{1}\cos \psi ^{\prime }+\left(
t_{3}-\chi _{3}\sin \psi ^{\prime }\right) \sigma _{3}\right]   \nonumber \\
&=&\left( 
\begin{array}{cc}
\frac{2}{3}-\frac{1}{6}\sin \psi ^{\prime } & \frac{1}{2\sqrt{3}}\cos \psi
^{\prime } \\ 
\frac{1}{2\sqrt{3}}\cos \psi ^{\prime } & \frac{1}{3}+\frac{1}{6}\sin \psi
^{\prime }%
\end{array}%
\right) .  \label{eq46}
\end{eqnarray}%
The eigenvalues of $\rho _{2}^{\prime }$ are 
\begin{equation}
\beta _{1,2}^{s}=\frac{1}{2}\pm \frac{1}{6}\sqrt{4-2\sin \psi ^{\prime
}-2\sin ^{2}\psi ^{\prime }}.  \label{eq47}
\end{equation}%
When $p_{1}=p_{2}$ we have%
\begin{eqnarray}
\varrho ^{s} &=&\sum_{j}p_{j}\varepsilon ^{s}\left( \rho _{j}\right)  
\nonumber \\
&=&\left( 
\begin{array}{cc}
\frac{4}{6}-\frac{1}{12}\sin \psi ^{\prime } & \frac{\sqrt{3}}{12}\left(
1+\cos \psi ^{\prime }\right)  \\ 
\frac{\sqrt{3}}{12}\left( 1+\cos \psi ^{\prime }\right)  & \frac{2}{6}+\frac{%
1}{12}\sin \psi ^{\prime }%
\end{array}%
\right) .  \label{eq48}
\end{eqnarray}%
The eigenvalues of Eq.(\ref{eq48}) are%
\begin{equation}
\gamma _{1,2}^{s}=\frac{6\pm \sqrt{10-4\sin \psi ^{\prime }-2\sin ^{2}\psi
^{\prime }+6\cos \psi ^{\prime }}}{12}.  \label{eq49}
\end{equation}%
Thus, the capacity can be obtained as 
\begin{equation}
C_{pe}^{s}=\max_{\psi ^{\prime }}\left\{ H\left( A\right) -\frac{1}{2}\left[
H\left( B\right) +H\left( \frac{1}{6}\right) \right] \right\} .  \label{eq50}
\end{equation}%
where $A=\left( 6-\sqrt{9-4\sin \psi ^{\prime }+\cos 2\psi ^{\prime }+6\cos
\psi ^{\prime }}\right) /12,$ $B=\left( 3-\sqrt{3-2\sin \psi ^{\prime }+\cos
2\psi ^{\prime }}\right) /6.$ The numerical work shows that if the input
states are orthogonal, namely, take $\psi ^{\prime }=3.14159red,$ the
information output is $I_{pe}^{s}=0.268277bits$ (which is bigger than that
obtained in \cite{Fuchs97} a little bit). This is not the maximum output
information. When the input states have angle $\psi ^{\prime }=3.20359red$
(here, $\psi ^{\prime }=\pi $ is the orthogonal case) we can obtain the
maximum output information, namely the capacity, $C_{pe}^{s}=0.268673bits.$
This is also coincided with the result of \cite{Fuchs97}, qualitatively, but
the quantity of capacity is less than that Fuchs' result a little bit. By
our method we also complete the demonstration that the splaying channel's
classical information capacity need not be achievable by orthogonal states. 

\section{Conclusions}

In this paper we have investigated the classical information capacities $%
C_{pe}$ for some well-known quantum noisy channels by using different
representations of qubit quantum states. It is shown that directly
calculating capacity with solving eigenvalues of output states is very
convenient but it only adapt to a few of channels. By using this method we
investigate the classical information capacities $C_{pe}$ of depolarizing
channel and erasure channel. We use of Bloch sphere representation of qubit
quantum states calculating the capacities of phase damping cannel, two-Pauli
cannel and flip channels. It shows that the Bloch sphere representation is
convenient for analytically calculating the classical information capacities
of some quantum noisy channels expressed by unital maps. We use of the
Stokes parametrization representation investigating the classical
information capacities of non-unital amplitude damping channel and splaying
channel. To the former we have obtained a analytical result which has not
been reported in other where to our knowledge, and to the latter our result
is coincided with Fuchs' original result qualitatively.

\begin{acknowledgments}
This project partly supported by Scientific Research Fund of Hunan
Provincial Education Department under Grand No. 01C036
\end{acknowledgments}


\begin{thebibliography}{99}
\bibitem{Holevo/9809023} A. S. Holevo, e-print quant-ph/9809023.

\bibitem{NandCbook} M. A. Nielsen and I. L. Chuang, 2000 \emph{Quantum
Computation and Quantum Information}, Cambridge Press.

\bibitem{Kholevo73} A. S. Kholevo, Probl. Inf. Transm.\ 9 (1973) 177.

\bibitem{Fuchs96} C. A. Fuchs, 1996 Ph.D thesis, University of New Mexico;
e-print quant-ph/9601020.

\bibitem{Holevo98} A. S. Holevo, IEEE Trans. Inf. Theory 44 (1998) 269;
e-print quant-ph/9611023.

\bibitem{Hausladen96} P. Hausladen, R. Jozsa, B. Schumacher, M.
Westmoreland, and W. K. Wootters, Phys. Rev. A 54 (1996) 1869.

\bibitem{Schumacher97} B. Schumacher, and M. D. Westmoreland, Phys. Rev. A
56 (1997) 131.

\bibitem{Holevo0106075} A. S. Holevo, e-print quant-ph/0106075.

\bibitem{Baruum9702049} H. Baruum, M. A. Nielsen, and B. Schumacher, e-print
quant-ph/9702049.

\bibitem{Schumacher96} B. Schumacher, and M. A. Nielsen, Phys. Rev. A 54
(1996) 2629.

\bibitem{Bennett99} C. H. Bennett, P. W. Shor, J. A. Smolin, and A. V.
Thapliyal, Phys. Rev. Lett. 83 (1999) 3081.

\bibitem{LF02} X. T. Liang, H. Y. Fan, Mod. Phys. Lett. B 16 (2002) 441.

\bibitem{Uhlmann01} A. Uhlmann, J. Phys. A: math. Gen. 34 (2001) 7047.

\bibitem{Liang02} X. T. Liang, Mod. Phys. Lett. B 16 (2002) 19.

\bibitem{King01} C. King and M. B. Ruskai, IEEE Trans. Info. Theory 47
(2001) 192; e-print quant-ph/9911079.

\bibitem{GBP97} M. Grassl, T. Beth, and T. Pellizzari, Phys. Rev. A 56
(1997) 33.

\bibitem{BDS97} C. Bennett, D. P. Divincenzo, and Smolin, Phys. Rev. Lett.
78 (1997) 3217.

\bibitem{Keyl0202112} M. Keyl, e-print quant-ph/0202112.

\bibitem{Preskill} J. Preskill, \emph{Quantum Information and Computation},
http:// www. theory. caltech. edu/ \symbol{126}preskill/ ph229.

\bibitem{Fuchs97} C. Fuchs, Phys. Rev. Lett. 79 (1997) 1162.
\end{thebibliography}
\end{document}